\documentclass[twocolumn,amsmath,amssymb,prl,aps,superscriptaddress]{revtex4-1}
\usepackage[version=3]{mhchem} % Formula subscripts using \ce{}
\usepackage{amsmath}
\usepackage{upgreek}
\usepackage{color}
\usepackage{natbib}
\usepackage{graphicx,subfigure}% Include figure files

\usepackage{xspace}

\begin{document}

\title{Curvature-guided motility of microalgae in geometric confinement}

\author{Tanya Ostapenko}
\affiliation{Max Planck Institute for Dynamics and Self-Organization (MPIDS), Am Fa{\ss}berg 17, D-37077 G\"{o}ttingen, Germany}

\author{Fabian Jan Schwarzendahl}
\affiliation{Max Planck Institute for Dynamics and Self-Organization (MPIDS), Am Fa{\ss}berg 17, D-37077 G\"{o}ttingen, Germany}

\author{Thomas B\"{o}ddeker}
\affiliation{Max Planck Institute for Dynamics and Self-Organization (MPIDS), Am Fa{\ss}berg 17, D-37077 G\"{o}ttingen, Germany}

\author{Christian Titus Kreis}
\affiliation{Max Planck Institute for Dynamics and Self-Organization (MPIDS), Am Fa{\ss}berg 17, D-37077 G\"{o}ttingen, Germany}

\author{Jan Cammann}
\affiliation{Max Planck Institute for Dynamics and Self-Organization (MPIDS), Am Fa{\ss}berg 17, D-37077 G\"{o}ttingen, Germany}

\author{Marco G. Mazza}
\affiliation{Max Planck Institute for Dynamics and Self-Organization (MPIDS), Am Fa{\ss}berg 17, D-37077 G\"{o}ttingen, Germany}

\author{Oliver B\"{a}umchen}
\thanks{To whom correspondence should be addressed. E-mail: oliver.baeumchen@ds.mpg.de}

\affiliation{Max Planck Institute for Dynamics and Self-Organization (MPIDS), Am Fa{\ss}berg 17, D-37077 G\"{o}ttingen, Germany}

%\date{\today}

%\setlength{\baselineskip}{1.3\baselineskip}

\begin{abstract} 
\textbf{Microorganisms, such as bacteria and microalgae, often live in habitats consisting of a liquid phase and a plethora of  interfaces. The precise ways in which these motile microbes behave in their confined environment remain unclear. Using experiments, Brownian dynamics simulations, and analytical theory, we study the motility of a single \textit{Chlamydomonas} microalga in an isolated microhabitat with controlled geometric properties. We demonstrate how the geometry of the habitat controls the cell's navigation in confinement. The probability of finding the cell swimming near the boundary scales linearly with the wall curvature, as seen for both circular and elliptical chambers. 
The theory, utilizing an asymmetric dumbbell model of the cell and steric wall interactions, captures this curvature-guided navigation quantitatively with no free parameters.}
\end{abstract}

\maketitle

%%%%%%%%%%%%%%%%%%%%%
\section*{Introduction}
%%%%%%%%%%%%%%%%%%%%%

% TOPIC: General introduction to life in complex geometries
Life in complex geometries manifests itself at the microscopic level through the myriad of ways in which microorganisms interact with their environment. This entails a broad spectrum of microbiological phenomena, ranging from amoebic crawling \cite{sun,wu} and fibroblast migration \cite{jeon2015} guided by nano-patterned substrates, the directional migration of epithelial cells on curved surfaces \cite{yevick} and microbial proliferation in space-limited environments \cite{delarue2016} to the motility of microbiological swimmers in confinement \cite{berg1990}. In fact, the natural habitats for microbial life are often non-bulk situations, including aqueous microdroplets \cite{meckenstock2014} and the interstitial space of porous media, such as rocks \cite{wierzchos2012,robinson2015} and soil \cite{ranjard2001}. The study of how self-propelled microorganisms in a liquid medium interact with their geometric boundaries finds application in physiology with regards to spermatozoa motility in the reproductive tract \cite{eisenbach2006, denissenko2012a,kantsler2014a,nosrati2015}, the motion of parasites in the vertebrate bloodstream \cite{heddergott2012}, and in microbiology in the context of biofilm formation \cite{watnick2000,hallNatRM2004,flemmingNatRM2010,mazzaJPD2016}. 

% TOPIC: Swimming dynamics; hydrodynamics vs. contact/steric interactions at an interface
This lends itself to inquiry over the mechanisms involved in the interplay between the microorganism and its confining domain. While walkers and crawlers achieve locomotion by momentum transfer to a substrate, i.e.\ friction, microorganisms swimming in fluid medium might undergo long-range hydrodynamic interactions, in addition to contact interactions \cite{lauga2009,brotto2013}. For these microswimmers, a distinction between ``puller''- and ``pusher''-type swimmers is required \cite{elgeti2015}, since the flow fields around the two classes entail fundamental differences \cite{drescher2011a,drescher2010a,guasto2010a,leptos2009a}. Though much work was reported on the onset of collective effects of microswimmers due to reorientation, or lack thereof, at an interface (e.g.\ \cite{rothschild,frymier1995}), there are few studies on how a single microswimmer cell interacts with a boundary in the absence of cell-cell interactions. At flat interfaces, the contact of a spermatozoon's flagellum with a surface tends to rotate it towards a boundary, thus preventing its escape from flat or weakly curved surfaces \cite{kantsler2013a}. However, for the puller-type microswimmer \textit{Chlamydomonas}, a microalga with two anterior flagella, steric interactions and multiple flagellar contacts were found responsible for its microscopic scattering off of a flat interface \cite{kantsler2013a}. Single scattering events of an individual \textit{Chlamydomonas} cell were also reported at convex interfaces, where two regimes emerge as the cell scatters off: an initial, contact force regime and a second, hydrodynamics-dominated regime \cite{contino2015a}. Beyond these details of the microscopic interactions at interfaces, the way in which the global swimming behavior of a single cell is affected by the geometry of a confining domain remains elusive.

% TOPIC: What we do!
In this article, we report on the motility of a single \textit{Chlamydomonas reinhardtii} cell in tailor-made microhabitats to elucidate the effects of geometric confinement. We find that the dominant attributes of the swimming statistics are the alga's spatial confinement, which limits its motion to its swimming plane, and the compartment's curved boundary in this plane. We study precisely a single isolated cell in order to exclude any cell-cell interactions or collective effects. Our experiments are in quantitative agreement with Brownian dynamics simulations and analytical theory, whose main ingredients are steric wall interactions and the alga's torque at the compartment interface during a finite interaction time. While a conclusive description of the microscopic details of wall interactions might remain debated today, our results illuminate how a single puller-type cell's navigation in confinement is primarily dominated by the details of the environment's geometric constraints.

%%%%%%%%%%%%%%%%%%%%%
\section*{Results}
%%%%%%%%%%%%%%%%%%%%%

We employed optical microscopy techniques to study the motility of an individual \textit{Chlamydomonas} cell contained within an isolated quasi-two-dimensional microfluidic compartment. Experiments were performed in circular compartments with radii ranging from $25$ to $500$\,$\mu$m, as well as elliptical chambers. The height of all compartments was approximately 20\,$\mu$m, about one cell diameter (body and flagella); thus, out-of-plane reorientations of the cell are inhibited. Each experiment with a single, isolated cell was repeated up to 10 times using different cells each time. Trajectories were extracted using image processing and particle tracking algorithms (see \textit{Materials and Methods}). Each cell trajectory corresponds to a single, independent experiment for a particular compartment size, with no cell-cell interactions or collective effects. The large amount of experimental data provides reliable statistics, which are quantitatively compared to Brownian dynamics simulations and analytical theory.

%------------------------------------------------%
\subsection{Motility in Circular Compartments}
%------------------------------------------------%

\begin{figure}[t]
\centering
	\includegraphics[scale=0.44]{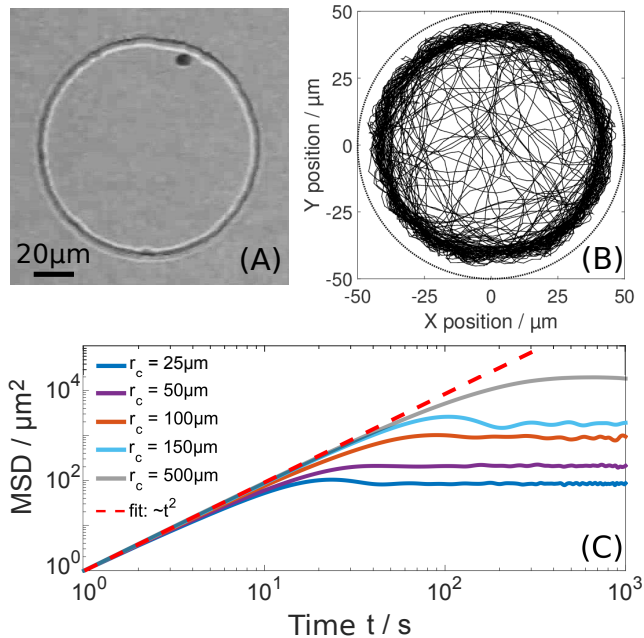}
\caption{Experimental design and trajectory analysis. (A) Optical micrograph of a single alga contained in a quasi-2D circular compartment. (B) Exemplar single-cell trajectory for $r_c = 50\,\mu$m. (C) Mean-squared displacements (solid lines) for different compartment radii. The dashed line is a $\sim t^{2}$ best fit to the short-time ballistic behavior.
}
\label{traj}
\end{figure}

% TOPIC: Particle trajectories
Figure~\ref{traj}A displays an image from a single experiment (see movie S1) for a compartment radius $r_c = 50\,\mu m$, from which the trajectory of the alga's body center was extracted (Fig.~\ref{traj}B). The alga's trajectory shows a higher density of trajectory points closer to the concave interface, as compared to the center of the compartment. We note that this higher density of points corresponds to an enhanced probability of finding the alga within the vicinity of the wall, and examine this quantity in detail during the course of our study.

% FIGURE 1

% FIGURE 2
\begin{figure*}
\centering
	\includegraphics[scale=0.35,width=\textwidth]{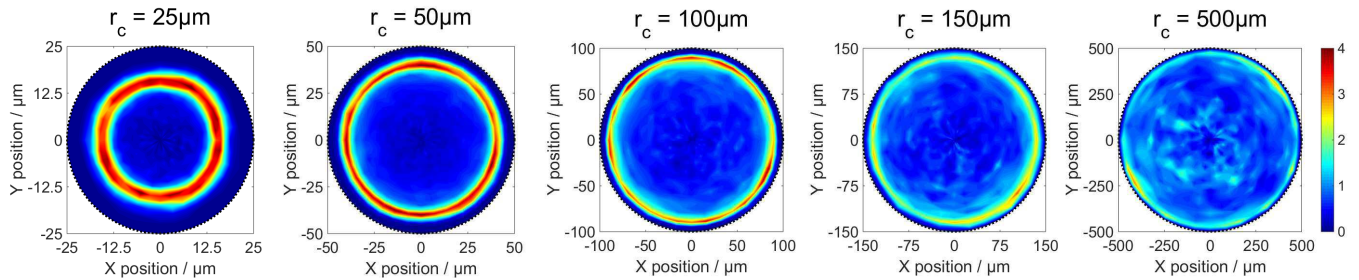}
\caption{Relative probability density for a single cell in circular confinement. Heat maps represent the alga's position obtained from experimental data for different compartment sizes: (L--R) $r_c = 25\,\mu$m, $50\,\mu$m, $100\,\mu$m, $150\,\mu$m, $500\,\mu$m. Each map contains statistically averaged data from a minimum of 2--5 independent experiments. For the applied normalization, see \textit{Materials and Methods}.}
\label{hm_pos}
\end{figure*}

% TOPIC: Mean-squared displacements
The mean-squared displacement (MSD) represents a parameter through which the type of swimming behavior that the alga exhibits is characterized. Here, the MSD for the observation time $t$ was extracted from a single alga's experimental trajectory for each compartment size; see Fig.~\ref{traj}C. An immediate observation is that the MSD curves show no clear transition between ballistic behavior, i.e.\ MSD $\sim t^2$, on short time scales to diffusive, i.e.\ MSD $\sim t$, on long time scales, as reported in previous studies on \textit{Chlamydomonas} swimming in unconfined 2D environments (transition time from ballistic to diffusive $\sim$\,2\,s) \cite{kantsler2013a}. A linear fit to the initial regime of the experimental data yields an exponent of 1.90$\pm$0.03, in agreement with a regime of ballistic swimming. The exponent for $r_c = 25\,\mu$m of 1.7$\pm$0.03 is attributed to the strong confinement situation, as the compartment is not more than approximately 2 times larger than the alga (body and flagella), representing the lower limit of our statistical analysis. On long time scales, the MSD reaches a plateau corresponding to the explorable area of its confined environment. Hence, we find that the alga's run-and-tumble-like motion in environments unconfined in the swimming plane \cite{polin2009a} becomes predominantly ballistic swimming in confinement. 

% TOPIC: relative probability densities
The experimental cell trajectories were statistically averaged and converted into relative probability density maps for each compartment size. Figure~\ref{hm_pos} displays a series of two-dimensional heat maps of the relative probability density of the cell's positions for different compartment sizes (see \textit{Materials and Methods} for details). Our experimental data provides evidence for a pronounced near-wall swimming effect inside the compartment, whose significance decreases for increasing compartment size.

\begin{figure}[t]
\centering
\includegraphics[scale=0.34]{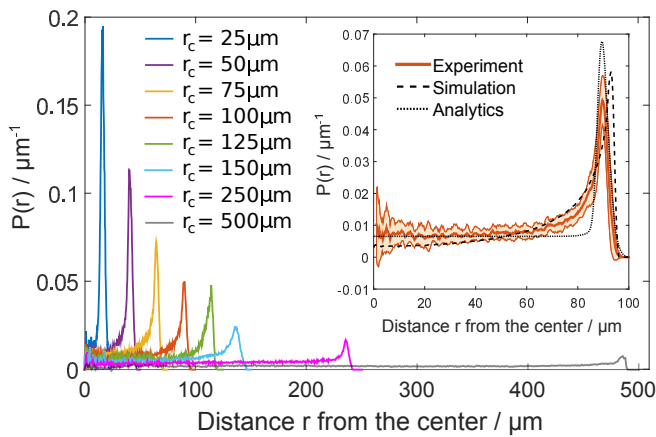}
\caption{Radial probability densities $P(r)$ for compartment sizes, $r_c$, between 25$\,\mu$m and 500$\,\mu$m. Each curve represents statistically averaged distributions from 2--10 independent measurements. As $r_c$ increases, the maximum of $P(r)$ close to the compartment wall decreases monotonously. The inset displays a close-up of the experimental data (7 independent experiments, solid line), Brownian dynamics simulation (dashed line) and analytical solution (dotted line) for $r_c$ = 100$\,\mu$m. The shaded background displays the standard deviation of the experimental data. See \textit{Materials and Methods} for details.}
\label{rpd_collapsed}
\end{figure}

% EQN for P(r)
This near-wall swimming effect is further quantified by azimuthally collapsing the heat maps from Fig.~\ref{hm_pos} into radial probability densities, $P(r)$, as depicted in Fig.~\ref{rpd_collapsed}. We define $P(r)$ as:
\begin{equation}
P(r) = \frac{h(r) / 2 \pi r \Delta r}{\int_{0}^{r_c} \frac{h(r)}{2 \pi r \Delta r}dr}
\label{pr}
\end{equation}
\\
where $r$ is the distance from the center of the compartment, and $h(r)$ is the count of all the alga's positions in a circular shell at distance $r$ with thickness $\Delta r$. In order to compare data from different compartment sizes, we normalize $P(r)$ such that $\int_{0}^{r_c} P(r) dr = 1$. Note that a homogeneous distribution of trajectory points would result in $P(r) = 1 / r_c = const.$ by this definition. In Fig.~\ref{rpd_collapsed}, particularly the inset, we observe that $P(r)$ starts from a plateau in proximity of the compartment's center and increases significantly close to the wall. The lateral extent (full-width-half-maximum) of the peak of $P(r)$ ranges from 3--5\,$\mu$m, about half a cell body diameter; the peak position is consistently 9--11\,$\mu$m away from the wall. At the compartment wall, $P(r)$ drops off, representing a possible zone of flagella-wall contact interactions. The maximum of $P(r)$ decreases for increasing compartment size from $r_c = 25\,\mu$m to $r_c = 500\,\mu$m, while the overall shape of $P(r)$ described above is preserved.

We also note that towards the compartment's center, where there is no significant influence of the wall on the swimming behavior, the alga swims with a typical velocity of 100$\pm10\,\mu$m/s. This is in agreement with swimming velocities reported in bulk \cite{polin2009a}. As a consequence of alga-wall interactions, the measured velocity slows down to 60$\pm20\,\mu$m/s within the near-wall swimming zone. \\

% FIGURE 3

%------------------------------------------------%
\subsection{Brownian Dynamics Simulations}
%------------------------------------------------%

\begin{figure*}[t]
        \centering
        \includegraphics[scale=0.35,width=\textwidth]{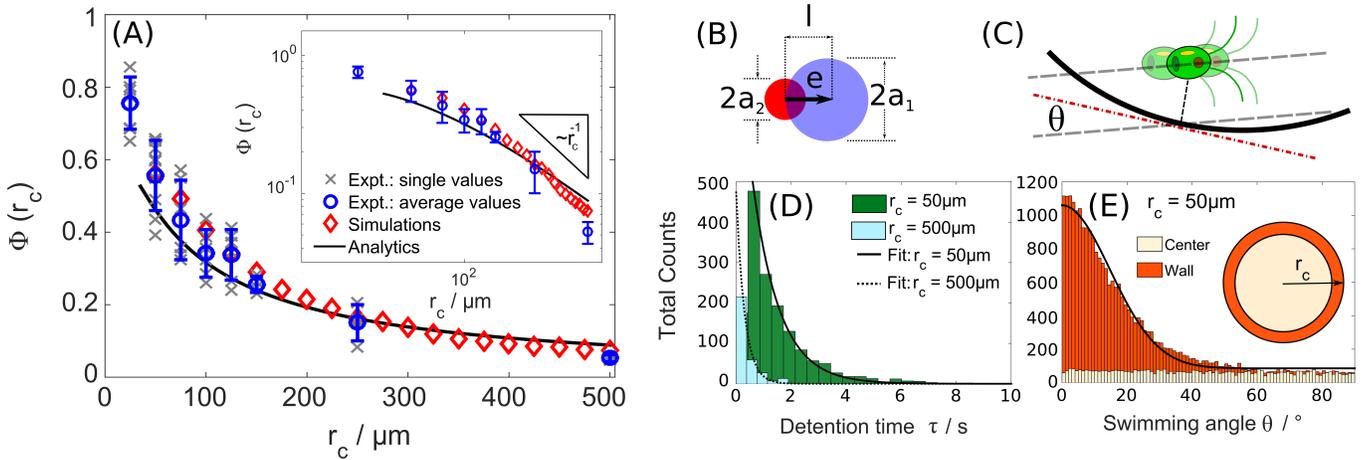}
        \caption{Statistics of near-wall swimming. (A) Near-wall swimming probability $\Phi (r_c)$ for compartment radii ranging from 25\,$\mu$m to 500\,$\mu$m: experimental data (circles) denote mean values averaged over multiple independent experiments (crosses), Brownian dynamics simulations (diamonds) and analytical calculation (solid line). (Inset) A log-log representation of the same data suggests a $1/r_{c}$ scaling for sufficiently large compartments, i.e.\ $r_c \geq 100\,\mu$m. (B) Asymmetric dumbbell model representing the \textit{Chlamydomonas} alga, see \textit{Materials and Methods} for details. (C) Schematic illustration of the swimming angle with respect to the wall tangential. (D) Detention time distributions for swimming near the concave interface for $r_c = 50\,\mu$m and $r_c = 500\,\mu$m. (E) Experimental swimming angle distributions for $r_c = 50\,\mu$m near and far (``center'') from the wall of the compartment. The inset illustrates the definition of ``wall'' and ``center'' for the purposes of the distributions.}
\label{fig:combined}
\end{figure*}

%TOPIC: Brownian dynamics simulations
We compared these experimental results to Brownian dynamics simulations adapted as follows. The \textit{Chlamydomonas} cell is modeled as an active asymmetric dumbbell consisting of two rigid spheres, inspired by the model developed in \cite{wysocki2015a}. The smaller sphere represents the cell's body and the larger sphere mimics the stroke-averaged area covered by the beating of the two anterior flagella (see Fig.~\ref{fig:combined}B). An asymmetric dumbbell is the simplest model for a cell to experience a torque during an interaction event with the confining wall. This torque is a major ingredient in our simulations, since it may reorient the alga away from the interface. All geometric and dynamic parameters that entered the simulations were either measured directly from our experiments or extracted from the literature. This includes a finite microscopic interaction time at the interface, which was inferred from image sequences in \cite{kantsler2013a}. In the simulations, we also consider stochastic noise to account for the biological nature of our microswimmer. However, the comparison to the analytical approach (provided in the \textit{SM}) suggests that this noise represents only a minor contribution. Hydrodynamic interactions are entirely absent in this model and the dynamics are solely determined by the repulsive force and torques at the confining wall (see \textit{Materials and Methods}).

The radial probability densities $P(r)$ were extracted from both simulations (see Fig.~S1 for heat maps) and analytical theory; exemplar curves are presented in the inset of Fig.~\ref{rpd_collapsed}. In general, we find good quantitative agreement of these data with the experiment, namely an increased likelihood of finding the alga located near the interface for small compartments.

% FIGURE 4

%------------------------------------------------%
\subsection{Near-Wall Swimming Probability}
%------------------------------------------------%

% TOPIC: peak ratio
We define the near-wall swimming probability, $\Phi (r_c)$, as the relative probability of finding the alga/dumbbell towards the wall as compared to the center. In our notation, this is written as:
% EQN for Phi(r_c)
\begin{equation}
\Phi (r_c) = 1 - \frac{r_c}{r_c - b} \int_{0}^{r_c - b} P(r) dr
\label{phi}
\end{equation}
\\
where $b$ is the extent of wall influence, measured from experiments as approximately 15\,$\mu$m and independent of compartment size. This distance $b$ defines the near-wall swimming region used later in our study.

Figure~\ref{fig:combined}A presents $\Phi (r_c)$ for experiments, Brownian dynamics simulations, and analytical theory, which all agree quantitatively and show a monotonic decrease for increasing compartment radius $r_{c}$. Furthermore, the data suggest a linear scaling of $\Phi (r_c)$ with the wall curvature, i.e.\ $1/r_{c}$ (see Fig.~\ref{fig:combined}A inset). The deviation from this linear scaling for strong confinement ($r_c < 100\,\mu$m) is consistent with the fact that the alga experiences a torque upon wall interaction that may allow it to occasionally escape, even for highly curved walls. 

%------------------------------------------------%
\subsection{Temporal and Angular Statistics of Near-wall Swimming}
%------------------------------------------------%

%TOPIC: detention time
We define the wall detention time $\tau$ as the time that the alga spends within the near-wall swimming zone (distance \textit{b} from the wall). As depicted in Fig.~\ref{fig:combined}D, the distribution of experimental detention times exhibits a maximum at approximately $\tau_\mathrm{max}$ = 0.5\,s for $r_c = 50\,\mu$m, featuring an exponential decay $e^{-t / \tau^{\ast}}$ towards longer detention times with a decay time of $\tau^{\ast}$ = 1\,s. For larger compartments, the maxima of the detention time distributions are shifted to shorter times and the decay times are also significantly reduced (see Fig.~\ref{fig:combined}D): $\tau_\mathrm{max}$ $<$ 0.2\,s and $\tau^{\ast}$ = 0.3\,s are found for $r_c = 500\,\mu$m, indicating that the wall detention time becomes comparable to the typical interaction time reported for flat interfaces \cite{kantsler2013a}. 

Since the cell has a non-zero velocity at all times during experiments, these results imply that the alga tends to spend more time swimming near the interface in smaller compartments than for larger ones. However, the alga's detention time at the interface does not yield any information about the directionality of the alga's swimming. It also does not allow us to distinguish between two possible extreme cases: (i) the alga probes the wall repeatedly at the same location, eventually escaping the interface after some time, and (ii) the alga swims non-stop either clock- or counterclockwise along the curved interface.

%TOPIC: swimming angle
In order to characterize the swimming direction, we analyzed the local swimming angle $\theta$ (see Fig.~\ref{fig:combined}C), which is measured relative to the local wall tangent. We consider two regions in the compartment: ``wall" and ``center" (see Fig.~\ref{fig:combined}E inset). As discussed above, the near-wall region is defined from a distance $b$ from the interface for all compartment sizes. All trajectories outside this region contribute to the swimming angle distribution towards the center of the compartment. As shown in Figure~\ref{fig:combined}E, we find that $\theta$ displays an isotropic distribution towards the center of the compartment. However, within the near-wall swimming zone, $\theta$ shows a maximum around zero swimming angle, indicating that the wall induces a preferred swimming direction parallel to the concave interface. The distribution decays towards larger angles in line with Gaussian statistics (standard deviation 12$^{\circ}$, solid line in Fig.~\ref{fig:combined}E). Note that these experimental distributions represent all navigational movement of the alga, which may include any microscopic interactions the alga might have with the compartment boundaries.

% TOPIC: Microscopic explanation of how the alga swims in circular compartments
Based on the aforementioned analysis, the alga undergoes the following process in isolated microcompartments: Upon interaction with an interface, the alga reorients due to its characteristic torque and scatters off at some shallow angle (see also \cite{kantsler2013a}). It subsequently continues swimming predominantly ballistically in the compartment. If the compartment is sufficiently curved, the alga will encounter another section of the interface in a short time, interact, scatter off, and continue swimming. This process will repeat itself such that for small compartments (high curvature), it appears that the alga swims non-stop parallel to the interface (see movie S1), since the alga will encounter another interface during its characteristic persistent swimming time. In contrast, for large compartments (low curvature), the alga will travel farther before meeting another interface. Thus, it is more likely that the alga's reorientation will direct it towards the compartment center. Nonetheless, due to the confinement the alga will encounter an interface before undergoing any run-and-tumble-like motion. Our Brownian dynamics simulations and analytical theory capture this process: using an asymmetric dumbbell model, the alga will naturally experience a torque at the interface and reorient with a finite interaction time, subsequently encountering another interface before it can ``tumble''.

%------------------------------------------------%
\subsection{Geometries with Convex Interfaces}
%------------------------------------------------%

%TOPIC: Pillars
The results described above suggest that the alga may not remain swimming near the convex interface of a cylindrical obstacle, in contrast to pusher-type swimmers, e.g.\ bacteria \cite{spagnolie2014a} and spermatozoa \cite{takagi2014a}. An alga contained in a circular compartment including a central pillar establishes a simultaneous comparison between the behavior at concave and convex interfaces within the same compartment. The analysis of experimental and simulated trajectories (see Fig.~S2) show that the alga scatters off at the pillar and escapes the convex wall. This observation is consistent with studies on single microscopic scattering events at flat \cite{kantsler2013a} and convex \cite{contino2015a} interfaces. We also note that $P(r)$ in the vicinity of the convex pillar interface is slightly increased as compared to the value of $P(r)$ away from any walls (see Fig.~S2), which can be attributed to the alga's finite interaction time at the interface.

%------------------------------------------------%
\subsection{Motility in Elliptical Compartments}
%------------------------------------------------%
% FIGURE 5
\begin{figure}[h]
\centering
   \includegraphics[scale=0.2,width=\linewidth]{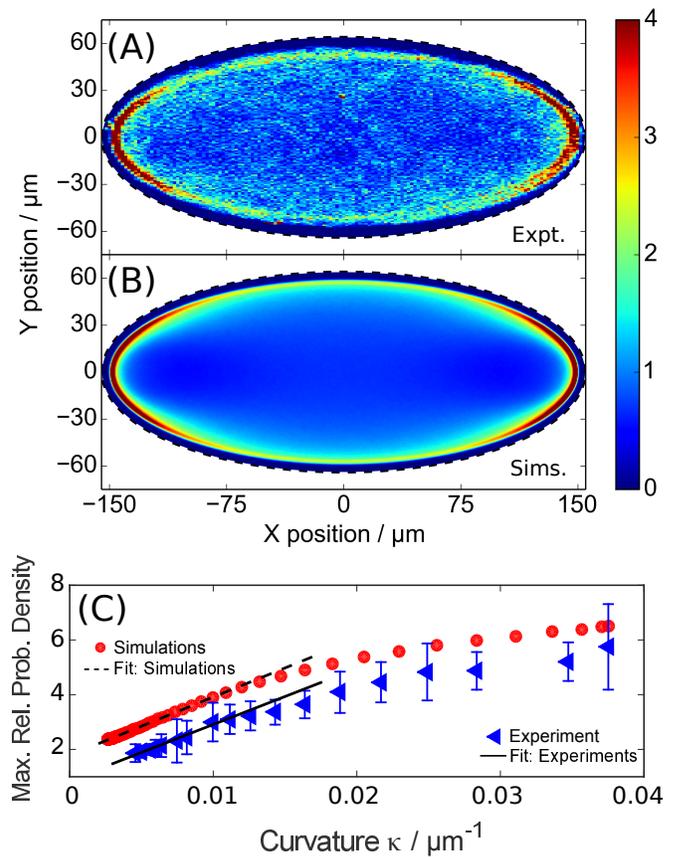}
\caption{Motility in elliptical compartments. Relative probability density heat maps extracted from experiments (A) and Brownian dynamics simulations (B) both display an enhanced probability density in the apex regions of the elliptical compartment (eccentricity = 0.91). (C) The local maximum relative probability density in both experiments and simulations scales with the local wall curvature $\kappa$. Lines represent linear fits to the experimental (straight) and simulation (dashed) data for wall curvatures $\kappa \leq 0.01 \mu$m$^{-1}$.}
\label{convex}
\end{figure}

% TOPIC: Elliptical Chambers
Finally, we consider elliptical chambers, which are compartments possessing a curvature gradient along its wall. Experiments (Fig.~\ref{convex}A) and Brownian dynamics simulations (Fig.~\ref{convex}B) show a higher likelihood of finding the alga in one of the apex regions of the compartments. We find in both experiments and simulations that the local maximum of the relative probability density in elliptical compartments scales linearly with the corresponding local wall curvature $\kappa$ (Fig.~\ref{convex}C), in line with the results obtained for circular compartments (see Fig.~\ref{fig:combined}A). 
We observe a deviation from this linear scaling in both experiments and simulations for higher wall curvatures, as also seen for circular compartments. Hence, we have established unambiguous evidence that the (local) wall curvature controls the near-wall swimming effect in confinement and that the results described for circular compartments cannot be attributed to a trivial surface-to-volume effect (see \cite{fily2014a}). Note that a torque-free, spherical active Brownian particle cannot reproduce the experimental data (see Fig.~S4). Thus, we find that \textit{Chlamydomonas}' motility in confinement is governed by reorientations of the cell due to steric wall interactions and the cell's characteristic torque.

%%%%%%%%%%%%%%%%%%%%%
\section*{Discussion}
%%%%%%%%%%%%%%%%%%%%%

In the absence of external flow, cell-cell interactions, photo- and chemotaxes, we isolated a curvature-guided motility mechanism for a single microalga in a confined microfluidic habitat with controlled geometric properties. The concave nature of the confining walls leads to an enhanced probability of near-wall swimming for puller-type microswimmers, as quantified by a statistical analysis of experimental cell trajectories. Brownian dynamics simulations based on an active asymmetric dumbbell model, as well as analytical theory, quantitatively capture the experiments and validate a characteristic curvature scaling of the near-wall swimming probability. The main ingredients of this curvature guidance are the torque that the alga experiences during an interaction event with the wall, the compartment's wall curvature, and the suppression of the alga's diffusive swimming regime in confinement. We highlight that hydrodynamics are not explicitly necessary to understand this swimming behavior, yet they might be required to capture the microscopic details of flagellar interactions with interfaces, as previously reported \cite{kantsler2013a,contino2015a}. Our results suggest that enhanced near-wall swimming in confinement is not exclusive to microorganisms propelling themselves by rear-mounted appendages, but is also manifested by microorganisms with anterior flagella, such a microalga.

These results pave the way towards a fundamental understanding of the motility of microorganisms in their natural habitats, which typically constitute a plethora of curved interfaces. \textit{Chlamydomonas} is a soil-dwelling microorganism and represents one of the prime model organisms to study cellular processes, e.g.\ ciliary functions in eukaryotes and photosynthesis, with a wide range of applications including the synthesis of therapeutic proteins and biofuels. In light of the idea that surface association represents an efficient way of harboring a favorable environment rather than being swept away by a liquid current \cite{watnick2000}, the observed swimming behavior might represent an advantage in the cell's natural habitat. 

From a technological perspective, we anticipate that these insights may inspire new design principles for the guidance of cellular motion, e.g.\ for the targeted delivery of (pharmacological) cargo \cite{weibel2005} and for directing motile eukaryotic cells through microfluidic channels in bioengineering applications \cite{ai2012,min2014,das2015}, complementary to existing rectification approaches \cite{kantsler2013a}. Moreover, an immediate consequence of enhanced detention times at a highly curved wall is a greater likelihood for the adhesion of planktonic cells at walls, which can trigger the colonization and formation of biofilms in liquid-immersed porous media. In such heterogeneous environments, biofilm formation has substantial implications on transport processes and permeability \cite{cunningham1991}. Thus, we expect that these insights are highly relevant in environmental applications related to bioremediation of contaminated groundwater and soil, water filtration systems, and bioreactors for the production of biofuels based on photosynthetic microalgae \cite{schenk2008a}.

%%%%%%%%%%%%%%%%%%%%%
\section*{Materials and Methods}
%%%%%%%%%%%%%%%%%%%%%

% TOPIC: Cell culturing
\subsection{Cell Cultivation} Cultures of wild-type \textit{Chlamydomonas reinhardtii} (SAG 11-32b) were cultivated axenically in Tris-Acetate-Phosphate (TAP) medium on a 12\,h--12\,h day-night cycle, with daytime temperature of 24\,$^{\circ}$C and nighttime temperature of 22\,$^{\circ}$C in a Memmert IPP 100Plus incubator. The daytime light intensity was held at 1000--2000 Lux, and reduced to 0\% during the night. All experiments were performed at the same time in the cell's life cycle in order to ensure consistency in the cell's size and behavior. In preparation for experiments, 15\,mL of cell suspension was centrifuged for 10 minutes at 100\,$g$ at room temperature; 10-13\,mL of solution was subsequently removed and the remaining 2-5\,mL suspension was allowed to relax for 60-90 minutes. This suspension was diluted with room temperature TAP to enhance the likelihood for capturing precisely one cell in an isolated microcompartment.

% TOPIC: Microfluidics
\subsection{Microfluidics} Arrays of stand-alone (i.e.\ no inlet or outlet) circular microfluidic compartments with a height of 20\,$\mu$m were created using standard PDMS-based soft lithography techniques in a cleanroom. Additional experiments were conducted with circular compartments containing a pillar located at the center of the compartment, as well as elliptical chambers. Prior to experiments, both the PDMS device and a glass microscope slide were cleaned using air plasma (Electronic Diener Pico plasma system, 100\% exposure, 30\, seconds). After plasma cleaning, a small amount of 8wt\% polyethelyne glycol was gently rinsed over both surfaces to prevent adhesion of the alga to the surfaces. After placing a droplet of the diluted algal suspension onto the feature side of the PDMS, the glass slide was placed on top and gently pressed to seal the compartment. Only compartments containing precisely one cell were used for experiments.

% TOPIC: Microscopy
\subsection{Microscopy} Cell imaging was conducted using an Olympus IX-81 inverted microscope contained in a closed box on a passive anti-vibration table with an interference bandpass filter ($\lambda \geq$ 671\,nm, full-width-half-maximum of 10\,nm) in order to avoid any photoactive response of the cell. Videos ranging from 5-30\,minutes were recorded using a Canon 600D camera at a frame rate of 24 frames per second at full resolution (1920\,pxl$\times$1080\,pxl). This corresponds to approximately 7$\times10^{3}$--40$\times10^{3}$ total trajectory points for a single experiment. The single-cell experiments were repeated 2--10 times for each compartment size, corresponding to 35$\times10^{3}$--120$\times10^{3}$ total trajectory points.

% TOPIC: Image Processing and Particle Tracking
\subsection{Image Processing and Particle Tracking} The videos were sequenced into 8-bit grayscale images with improved contrast using custom-made \textsc{Matlab} algorithms. The compartment boundaries were manually defined in order to denote the region of interest, as well as the compartment's center, for particle tracking. Two-dimensional particle detection was performed using algorithms written for colloidal systems; particle tracking was subsequently done in \textsc{Matlab} based on tracking algorithms developed by Crocker and Grier \cite{crocker1996a}.

% TOPIC: Data Analysis
\subsection{Data Analysis} Mean-squared displacements (MSD) were extracted using a custom-made \textsc{Matlab} script, based on trajectories containing a minimum of 7$\times10^{3}$--14$\times10^{3}$ data points for the $r_c = 25-250\,\mu$m and at least 40$\times10^{3}$ data points for the $r_c = 500\,\mu$m. This corresponds to MSD curves containing 2$\times10^{3}$--10$\times10^{3}$ points each.

A custom-made \textsc{Matlab} algorithm based on a pixel grouping method for the data binning was applied to the trajectory data to compute the relative probability density heat maps by $c(x,y) = n_\mathrm{bin}/(A_\mathrm{frac}\Sigma n_\mathrm{bin})$, where $n_\mathrm{bin}$ is the number of trajectory points within bin area $A_\mathrm{bin}$, and $A_\mathrm{frac} = A_\mathrm{bin}/A_\mathrm{chamber}$ is a geometric normalization factor. The radial probability densities $P(r)$ of the alga's distance $r$ from the center of the compartment were calculated using Eqn.~\ref{pr} such that $\int_{0}^{r_c} P(r)\mathrm{d}r = 1$. We define the near-wall swimming probability as in Eqn.~\ref{phi}.

% TOPIC: Simulations
\subsection{Brownian Dynamics Simulations: Details of Steric Wall Interactions}
The force acting on the dumbbell at the wall is given by $\vec{F}_w = \vec{F}_1 + \vec{F}_2$ with $\vec{F}_{\alpha} = - \vec{\nabla} U_{\alpha}(r)$, $\alpha=1,2$. We use the Weeks--Chandler--Anderson repulsive potential
$U_{\alpha} (d)/(k_\mathrm{B}T) =  4 \epsilon \left[ \left(\frac{a_{\alpha}}{d} \right)^{12}-  \left( \frac{a_{\alpha}}{d}\right)^6 \right] + \epsilon$, if $d < 2^{1/6}a_{\alpha}$, and   $0$ otherwise,                   
where $d$ is the distance of the sphere $\alpha\in \{1,2\}$ to the wall of the compartment, $a_{1}=5\,\mu$m, $a_2=2.5\,\mu$m are the radii of the spheres (see Fig.~\ref{fig:combined}B) and $\epsilon=10$ is chosen to achieve a strong screening. Furthermore, the torque is given by $\vec{T}_w = \vec{T}_1 + \vec{T}_2$ where 
$\vec{T}_1 = (\vec{r}_1 - \vec{r}) \times \vec{F}_1 = l (\vec{e} \times \vec{F_1})/2$,  $\vec{T}_2 = -l (\vec{e} \times \vec{F_2})/2$, and $l=5\,\mu$m.
The position $\vec{r}$ of the dumbbell's center of mass has the following equation of motion
$\frac{ \text{d} \vec{r} }{ \text{d} t } =  v_0 \vec{e} + \gamma_w \vec{F}_w + \vec{\eta}$.
Here, $\vec{F}_w $ is the steric wall interaction 
and $\vec{\eta}$ is a Gaussian white noise with zero mean and $\langle \vec{\eta}(t) \vec{\eta}(t') \rangle = 2 k_\mathrm{B}T\gamma_w \mathbf{1} \delta(t-t')$.
We use $v_0 = 60\,\mu \text{m/s}$ and $k_\mathrm{B}T\gamma_w = 20\,\mu$m$^{2}/s$, both based on experimental measurements.
Furthermore, the cell swims in the direction $\vec{e}$ represented by a versor (a unit vector) pointing from the second to the first sphere (see Fig.~\ref{fig:combined}B). The orientational equation of motion is
$\frac{ \text{d} \vec{e} }{ \text{d} t } =  (  \vec{T}_w / \tau_w + \vec{\xi}) \times  \vec{e}$,
where $\vec{T}_w$ is the torque acting at the wall 
and $\vec{\xi}$ is a Gaussian white noise with $\langle \vec{\xi}(t) \vec{\xi}(t') \rangle = \frac{2k_\mathrm{B}T}{ \tau_p } \mathbf{1} \delta(t-t')$, representing the tumble motion of the cell.  
Note that the shear time at the wall  $\tau_w = 0.15\,\text{s}$ (extracted from \cite{kantsler2013a}) 
and the persistence time $\tau_p = 5.1\,\text{s}$ (taken from \cite{polin2009a}) of synchronous flagella beating are not connected \emph{via} the fluctuation-dissipation theorem. 
This is motivated by the fact that the tumble time is associated with the active motion of the cell, whereas the shear time is connected to the interactions between wall and cell.
Our simulated system quantitatively reproduces the maximum of the microscopic scattering angle distribution, found experimentally in \cite{kantsler2013a} (see Fig.\ S3).

%\end

%%%%%%%%%%%%%%%%%%%%%
\section*{Acknowledgements}
%%%%%%%%%%%%%%%%%%%%%

\textbf{General:} The authors gratefully acknowledge M.\ Lorenz and the Algae Culture Collection (SAG) in G\"{o}ttingen, Germany, for providing the \textit{Chlamydomonas reinhardtii} strain SAG 11-32b. We also thank S.\ Herminghaus for insightful discussions, as well as A.\ Schella and D.\ Lavrentovich for discussions and technical assistance. \textbf{Funding:} F.S and M.G.M. acknowledge financial support from the DFG Collaborative Research Center SFB 937 (Project A20). O.B. acknowledges support from the ESPCI Joliot Chair. \textbf{Author contributions:} M.G.M. and O.B. designed research; T.O., F.S., T.B., C.K., J.C., M.G.M. and O.B. performed research; T.O., F.S., T.B., C.K., J.C., M.G.M. and O.B. analyzed data; T.O., F.S., M.G.M and O.B. wrote the paper. \textbf{Competing interests:} The authors declare that they have no competing interests. \textbf{Data and materials availability:} All data needed to evaluate the conclusions in the paper are present in the paper and/or the Supplementary Materials. Additional data related to this paper may be requested from the authors.


\begin{thebibliography}{69}
\bibitem[{Sun \emph{et~al.}(2015)}]{sun}
X. Sun, M.K. Driscoll, C. Guven, S. Das, C.A. Parent, J.T. Fourkas, W. Losert,
\newblock Asymmetric nanotopography biases cytoskeletal dynamics and promotes
  unidirectional cell guidance.
\newblock \emph{Proc. Natl Acad. Sci. USA} \textbf{112}, 12557--12562 (2015).

\bibitem[{Wu \emph{et~al.}(2015)}]{wu}
H. Wu, M. Thi\'{e}baud, W.-F. Hu, A. Farutin, S. Rafa\"{i}, M.-C. Lai, P. Peyla, C. Misbah,
\newblock Amoeboid motion in confined geometry.
\newblock \emph{Phys. Rev. E} \textbf{92}, 050701 (2015).

\bibitem[{Jeon \emph{et~al.}(2015)}]{jeon2015}
H. Jeon, S. Koo, W.M. Reese, P. Loskill, C.P. Grigoropoulos, K.E. Healy,
\newblock Directing cell migration and organization via nanocrater-patterned
  cell-repellent interfaces.
\newblock \emph{Nature Materials} \textbf{14}, 918--923 (2015).

\bibitem[{Yevick \emph{et~al.}(2015)Yevick, Duclos, Bonnet \&
  Silberzan}]{yevick}
H.~G. Yevick, G. Duclos, I. Bonnet, P. Silberzan,
\newblock Architecture and migration of an epithelium on a cylindrical wire.
\newblock \emph{Proc. Natl Acad. Sci. USA} \textbf{112}, 5944--5949 (2015).

\bibitem[{Delarue \emph{et~al.}(2016)}]{delarue2016}
M. Delarue, J. Hartung, C. Schreck, P. Gniewek, L. Hu, S. Herminghaus, O. Hallatschek,
\newblock Self-driven jamming in growing microbial populations.
\newblock \emph{Nature Physics} \textbf{12}, 762--766 (2016).

\bibitem[{Berg \& Turner(1990)}]{berg1990}
H.~C. Berg, L. Turner,
\newblock Chemotaxis of bacteria in glass capillary arrays. {E}scherichia coli,
  motility, microchannel plate, and light scattering.
\newblock \emph{Biophys. J.} \textbf{58}, 919--930 (1990).

\bibitem[{Meckenstock \emph{et~al.}(2014)}]{meckenstock2014}
R.~U. Meckenstock, F. von Netzer, C. Stumpp, T. Lueders, A.M. Himmelberg, N. Hertkorn, P. Schmitt-Kopplin, M. Harir, R. Hosein, S. Haque, D. Schulze-Makuch,
\newblock Water droplets in oil are microhabitats for microbial life.
\newblock \emph{Science} \textbf{345}, 673--676 (2014).

\bibitem[{Wierzchos \emph{et~al.}(2012)Wierzchos, de~los R{\'\i}os \&
  Ascaso}]{wierzchos2012}
J. Wierzchos, A. de~los R{\'\i}os, C. Ascaso,
\newblock Microorganisms in desert rocks: the edge of life on Earth.
\newblock \emph{Int. Microbiol.} \textbf{15}, 171--181 (2012).

\bibitem[{Robinson \emph{et~al.}(2015)}]{robinson2015}
C.~K. Robinson, J. Wierzchos, C. Black, A. Crits-Christoph, B. Ma, J. Ravel, C. Ascaso, O. Artieda, S. Valea, M. Rold\'{a}n, B. G\'{o}mez-Silva, J. DiRuggiero,
\newblock Microbial diversity and the presence of algae in halite endolithic
  communities are correlated to atmospheric moisture in the hyper-arid zone of
  the Atacama Desert.
\newblock \emph{Env. Microbiol.} \textbf{17}, 299--315 (2015).

\bibitem[{Ranjard \& Richaume(2001)}]{ranjard2001}
L. Ranjard, A. Richaume,
\newblock Quantitative and qualitative microscale distribution of bacteria in
  soil.
\newblock \emph{Research in Microbiology} \textbf{152}, 707 -- 716 (2001).

\bibitem[{Eisenbach \& Giojalas(2006)}]{eisenbach2006}
M. Eisenbach, L.~C. Giojalas,
\newblock Sperm guidance in mammals---an unpaved road to the egg.
\newblock \emph{Nature Reviews Molecular Cell Biology} \textbf{7}, 276--285
  (2006).

\bibitem[{Denissenko \emph{et~al.}(2012)Denissenko, Kantsler, Smith \&
  Kirkman-Brown}]{denissenko2012a}
P. Denissenko, V. Kantsler, D.~J. Smith, J. Kirkman-Brown,
\newblock Human spermatozoa migration in microchannels reveals
  boundary-following navigation.
\newblock \emph{Proc. Natl Acad. Sci. USA} \textbf{109}, 8007 (2012).

\bibitem[{Kantsler \emph{et~al.}(2014)Kantsler, Dunkel, Blayney \&
  Goldstein}]{kantsler2014a}
V. Kantsler, J. Dunkel, M. Blayney, R.~E. Goldstein,
\newblock Rheotaxis facilitates upstream navigation of mammalian sperm cells.
\newblock \emph{eLIFE} \textbf{3}, e02403 (2014).

\bibitem[{Nosrati \emph{et~al.}(2015)Nosrati, Driouchi, Yip \&
  Sinton}]{nosrati2015}
R. Nosrati, A. Driouchi, C.~M. Yip, D. Sinton,
\newblock Two-dimensional slither swimming of sperm within a micrometre of a
  surface.
\newblock \emph{Nature Communications} \textbf{6}, 8703 (2015).

\bibitem[{Heddergott \emph{et~al.}(2012)}]{heddergott2012}
N. Heddergott, T. Kr\"{u}ger, S.B. Babu, A. Wei, E. Stellamanns, S. Uppaluri, T. Pfohl, H. Stark, M. Engstler,
\newblock Trypanosome motion represents an adaptation to the crowded
  environment of the vertebrate bloodstream.
\newblock \emph{PLoS Pathog} \textbf{8}, 1--17 (2012).

\bibitem[{Watnick \& Kolter(2000)}]{watnick2000}
P. Watnick, R. Kolter,
\newblock Biofilm, City of Microbes.
\newblock \emph{Journal of Bacteriology} \textbf{182}, 2675--2679 (2000).

\bibitem[{Hall-Stoodley \emph{et~al.}(2004)Hall-Stoodley, Costerton \&
  Stoodley}]{hallNatRM2004}
L. Hall-Stoodley, J.~W. Costerton, P. Stoodley,
\newblock Bacterial biofilms: from the natural environment to infectious
  diseases.
\newblock \emph{Nat. Rev. Microbiol.} \textbf{2}, 95--108 (2004).

\bibitem[{Flemming \& Wingender(2010)}]{flemmingNatRM2010}
H.-C. Flemming, J. Wingender,
\newblock The biofilm matrix.
\newblock \emph{Nat. Rev. Microbiol.} \textbf{8}, 623--633 (2010).

\bibitem[{Mazza(2016)}]{mazzaJPD2016}
M.~G. Mazza,
\newblock The physics of biofilms--an introduction.
\newblock \emph{J. Phys. D: Appl. Phys.} \textbf{49}, 203001 (2016).

\bibitem[{Lauga \& Powers(2009)}]{lauga2009}
E. Lauga, T.~R. Powers,
\newblock The hydrodynamics of swimming microorganisms.
\newblock \emph{Rep. Prog. Phys.} \textbf{72}, 096601 (2009).

\bibitem[{Brotto \emph{et~al.}(2013)Brotto, Caussin, Lauga \&
  Bartolo}]{brotto2013}
T. Brotto, J.-B. Caussin, E. Lauga, D. Bartolo,
\newblock Hydrodynamics of confined active fluids.
\newblock \emph{Phys. Rev. Lett.} \textbf{110}, 038101 (2013).

\bibitem[{Elgeti \emph{et~al.}(2015)Elgeti, Winkler \& Gompper}]{elgeti2015}
J. Elgeti, R.~G. Winkler, G. Gompper,
\newblock Physics of microswimmers---single particle motion and collective
  behavior: a review.
\newblock \emph{Rep. Prog. Phys.} \textbf{78}, 056601 (2015).

\bibitem[{Drescher \emph{et~al.}(2011)Drescher, Dunkel, Cisneros, Ganguly \&
  Goldstein}]{drescher2011a}
K. Drescher, J. Dunkel, L.~H. Cisneros, S. Ganguly, R.~E. Goldstein,
\newblock Fluid dynamics and noise in bacterial cell--cell and cell--surface
  scattering.
\newblock \emph{Proc. Natl Acad. Sci. USA} \textbf{108}, 10940--10945 (2011).

\bibitem[{Drescher \emph{et~al.}(2010)Drescher, Goldstein, Michel, Polin \&
  Tuval}]{drescher2010a}
K. Drescher, R.~E. Goldstein, N. Michel, M. Polin, I. Tuval,
\newblock Direct measurement of the flow field around swimming microorganisms.
\newblock \emph{Phys. Rev. Lett.} \textbf{105}, 168101 (2010).

\bibitem[{Guasto \emph{et~al.}(2010)Guasto, Johnson \& Gollub}]{guasto2010a}
J.~S. Guasto, K.~A. Johnson, J.~P. Gollub,
\newblock Oscillatory flows induced by microorganisms swimming in two
  dimensions.
\newblock \emph{Phys. Rev. Lett.} \textbf{105}, 168102 (2010).

\bibitem[{Leptos \emph{et~al.}(2009)Leptos, Guasto, Gollub, Pesci \&
  Goldstein}]{leptos2009a}
K.~C. Leptos, J.~S. Guasto, J.~P. Gollub, A.~I. Pesci, R.~E. Goldstein,
\newblock Dynamics of enhanced tracer diffusion in suspensions of swimming
  eukaryotic microorganisms.
\newblock \emph{Phys. Rev. Lett.} \textbf{103}, 198103 (2009).

\bibitem[{Rothschild(1963)}]{rothschild}
Rothschild.
\newblock Non-random distribution of bull spermatozoa in a drop of sperm
  suspension.
\newblock \emph{Nature} \textbf{198}, 1221 (1963).

\bibitem[{Frymier \emph{et~al.}(1995)Frymier, Ford, Berg \&
  Cummings}]{frymier1995}
P.~D. Frymier, R.~M. Ford, H.~C. Berg, P.~T. Cummings,
\newblock Three-dimensional tracking of motile bacteria near a solid planar
  surface.
\newblock \emph{Proc. Natl Acad. Sci. USA} \textbf{92}, 6195--6199 (1995).

\bibitem[{Kantsler \emph{et~al.}(2013)Kantsler, Dunkel, Polin \&
  Goldstein}]{kantsler2013a}
V. Kantsler, J. Dunkel, M. Polin, R.~E. Goldstein,
\newblock Ciliary contact interactions dominate surface scattering of swimming
  eukaryotes.
\newblock \emph{Proc. Natl Acad. Sci. USA} \textbf{110}, 1187--1192 (2013).

\bibitem[{Contino \emph{et~al.}(2015)Contino, Lushi, Tuval, Kantsler \&
  Polin}]{contino2015a}
M. Contino, E. Lushi, I. Tuval, V. Kantsler, M. Polin,
\newblock Microalgae scatter off solid surfaces by hydrodynamic and contact
  forces.
\newblock \emph{Phys. Rev. Lett.} \textbf{115}, 258102 (2015).

\bibitem[{Polin \emph{et~al.}(2009)Polin, Tuval, Drescher, Gollub \&
  Goldstein}]{polin2009a}
M. Polin, I. Tuval, K. Drescher, J.~P. Gollub, R.~E. Goldstein,
\newblock Chlamydomonas swims with two ``gears'' in a eukaryotic version of
  run-and-tumble locomotion.
\newblock \emph{Science} \textbf{325}, 487--490 (2009).

\bibitem[{Wysocki \emph{et~al.}(2015)Wysocki, Elgeti \& Gompper}]{wysocki2015a}
A. Wysocki, J. Elgeti, G. Gompper,
\newblock Giant adsorption of microswimmers: Duality of shape asymmetry and
  wall curvature.
\newblock \emph{Phys. Rev. E} \textbf{91}, 050302(R) (2015).

\bibitem[{Spagnolie \emph{et~al.}(2015)Spagnolie, Moreno-Flores, Bartolo \&
  Lauga}]{spagnolie2014a}
S.~E. Spagnolie, G.~R. Moreno-Flores, D. Bartolo, E. Lauga,
\newblock Geometric capture and escape of a microswimmer colliding with an
  obstacle.
\newblock \emph{Soft Matter} \textbf{11}, 3396--3411 (2015).

\bibitem[{Takagi \emph{et~al.}(2014)Takagi, Palacci, Braunschweig, Shelley \&
  Zhang}]{takagi2014a}
D. Takagi, J. Palacci, A.~B. Braunschweig, M.~J. Shelley, J. Zhang,
\newblock Hydrodynamic capture of microswimmers into sphere-bound orbits.
\newblock \emph{Soft Matter} \textbf{10}, 1784--1789 (2014).

\bibitem[{fily \emph{et~al.}(2014)Fily, Baskaran \& Hagan}]{fily2014a}
Y. Fily, A. Baskaran, M.F. Hagan,
\newblock Dynamics of self-propelled particles under strong confinement.
\newblock \emph{Soft Matter} \textbf{8}, 3002--3009 (2014).

\bibitem[{Weibel \emph{et~al.}(2005)}]{weibel2005}
D.~B. Weibel, P. Garstecki, D. Ryan, W.R. DiLuzio, M. Mayer, J.E. Seto, G.M. Whitesides,
\newblock Microoxen: microorganisms to move microscale loads.
\newblock \emph{Proc. Natl Acad. Sci. USA} \textbf{102}, 11963--11967 (2005).

\bibitem[{Ai \emph{et~al.}(2012)}]{ai2012}
X. Ai, Q. Liang, M. Luo, K. Zhang, J. Pan, G. Luo,
\newblock Controlling gas/liquid exchange using microfluidics for real-time
  monitoring of flagellar length in living Chlamydomonas at the single-cell
  level.
\newblock \emph{Lab on a Chip} \textbf{12}, 4516--4522 (2012).

\bibitem[{Min \emph{et~al.}(2014)Min, Yoon, Joo, Sim \& Shin}]{min2014}
S.~K. Min, G.~H. Yoon, J.~H. Joo, S.~J. Sim, H.~S. Shin,
\newblock Mechanosensitive physiology of Chlamydomonas reinhardtii under direct
  membrane distortion.
\newblock \emph{Scientific Reports} \textbf{4}, 4675 (2014).

\bibitem[{Das \emph{et~al.}(2015)}]{das2015}
S. Das, A. Garg, A.I. Campbell, J. Howse, A. Sen, D. Velegol, R. Golestanian, S.J. Ebbens,
\newblock Boundaries can steer active Janus spheres.
\newblock \emph{Nature Communications} \textbf{6}, 8999 (2015).

\bibitem[{Cunningham \emph{et~al.}(1991)Cunningham, Characklis, Abedeen \&
  Crawford}]{cunningham1991}
A.~B. Cunningham, W.~G. Characklis, F. Abedeen, D. Crawford,
\newblock Influence of biofilm accumulation on porous media hydrodynamics.
\newblock \emph{Environmental Science \& Technology} \textbf{25}, 1305--1311
  (1991).

\bibitem[{Schenk \emph{et~al.}(2008)}]{schenk2008a}
P.~M. Schenk, S.R. Thomas-Hall, E. Stephens, U.C. Marx, J.H. Mussgnug, C. Posten, O. Kruse, B. Hankamer,
\newblock Second generation biofuels: high-efficiency microalgae for biodiesel
  production.
\newblock \emph{Bioenerg. Res.} \textbf{1}, 20--43 (2008).

\bibitem[{Crocker \& Grier(1996)}]{crocker1996a}
J.~C. Crocker, D.~G. Grier,
\newblock Methods of digital video microscopy for colloidal studies.
\newblock \emph{J. Colloid Interface Sci.} \textbf{179}, 298--310 (1996).

\end{thebibliography}
\end{document}